\documentstyle[12pt,epsfig]{article}
\makeatletter
\def\section{\@startsection {section}{1}{\z@}{-1cm plus-1ex
    minus-.2ex}{1.5ex plus.2ex}{\reset@font\normalsize\bf}}
\def\subsection{\@startsection{subsection}{2}{\z@}{-0.7cm plus-1ex
    minus-.2ex}{1.0ex plus.2ex}{\reset@font\normalsize\it}}
\def\subsubsection{\@startsection{subsubsection}{3}{\z@}{-0.7cm plus-1ex
    minus-.2ex}{1.0ex plus.2ex}{\reset@font\normalsize}}
\def\paragraph{\@startsection
     {paragraph}{4}{\z@}{3.25ex plus1ex minus.2ex}{-1em}{\reset@font
     \normalsize\bf}}
\def\subparagraph{\@startsection
     {subparagraph}{4}{\parindent}{3.25ex plus1ex minus
     .2ex}{-1em}{\reset@font\normalsize\bf}}

\ifcase \@ptsize
\or % mods for 11 pt
\or % mods for 12 pt
\fi
\advance\textheight by \topskip
\advance\footskip by 6mm

\newcount\dfttsuboption
\def\dfttnum#1{\def\@dfttnum{#1}}
\dfttnum{??/xx}
\def\dfttdraft#1{\dfttsuboption=1\def\draftname{DRAFT #1\ }%
\def\baselinestretch{1.4}}   % draft has increased baselineskip!

{ \newcount\hour \newcount\minutes \newcount\hourint%
  \minutes=\time \hour=\time \divide\hour by 60%
  \hourint=\hour   \multiply\hourint by -60  \advance\minutes by \hourint%
  \xdef\@time{\the\hour:\ifnum\minutes>9\else0\fi\the\minutes}%
}

\def\@maketitle{\newpage
 \null\begingroup\def\baselinestretch{1}
\raggedleft\normalsize DFTT 9/96 \\
\raggedleft\normalsize MPI-PhT/96-9 \\
\raggedleft\normalsize LU TP 96-7 \\
\raggedleft\normalsize February 26th, 1996 \\
 \vskip 0.8cm
 \begin{center}%
  {\Large \@title \par}%
  \vskip 1.5em
  {\normalsize
   \lineskip .5em
   \begin{tabular}[t]{c}\@author
   \end{tabular}\par}%
  \ifnum\dfttsuboption=0 \vskip 1em{\footnotesize \@date}\fi%
 \end{center}%
 \par\endgroup
 \vskip 1cm}

\newenvironment{summary}{\begin{quote}\begin{center}\bf\abstractname\par%
\end{center}\vskip 0.25em}{\end{quote}\vskip 2em}

\def\eref#1{(\ref{#1})}
\makeatother

\dfttdraft{3}
\newcommand{\be}{\begin{equation}}
\newcommand{\ee}{\end{equation}}
\newcommand{\ba}{\begin{eqnarray}}
\newcommand{\ea}{\end{eqnarray}}

\newcommand{\NF}{\cal{N}_{\kern -1.9pt f}}     %math mode only
\newcommand{\NC}{\cal{N}_{\kern -1.7pt c}}     %math mode only
\newcommand{\pT}{{p\kern -.2pt\lower 4pt\mbox{\tiny T}}}    %works?
\newcommand{\pL}{{p\kern -.2pt\lower 4pt\hbox{\tiny L}}}    %works?
\title{ \bf Common origin of  the shoulder
 in multiplicity distributions  and of oscillations
 in the factorial cumulants to factorial moments ratio
  \thanks{\it Work supported in part by M.U.R.S.T. (Italy) under grant 1995.} }

\author{A.\  GIOVANNINI$^1$ \thanks{E-mail: giovannini@to.infn.it}\ , \
 S.\  LUPIA$^2$ \thanks{E-mail: lupia@mppmu.mpg.de}\ , \
 R.\ UGOCCIONI$^3$ \thanks{E-mail: roberto@thep.lu.se} \\ \\
\it $^1$ Dip. Fisica Teorica and I.N.F.N. -- Sezione di Torino, \\
\it via Giuria 1, I-10125 Torino, Italy  \\  \\
\it $^2$  Max-Planck-Institut f\"ur Physik, Werner-Heisenberg-Institut \\
\it F\"ohringer Ring 6, D-80805 M\"unchen, Germany \\  \\
\it $^3$ Dept. of Theoretical Physics, University of Lund \\
\it S\"olvegatan 14 A, S 223 62, Lund, Sweden}

\begin{document}

\maketitle

\begin{summary}
The shoulder structure of
 charged particles multiplicity distributions (MD's) in full phase space
 in $e^+e^-$ annihilation at the $Z_0$ peak and the quasi-oscillatory behavior
 of the ratio of factorial cumulants over factorial
moments, $H_q$, as a function of the order $q$,
are quantitatively reproduced within a simple parametri\-zation of the MD
in terms of a weighted superposition of two  Negative Binomial
Distributions,
 associated to two- and multi-jet production, i.e., to hard gluons
radiation.
\end{summary}

\vspace{0.5cm}

\newpage

\section{Introduction}

A definite understanding of the
multiparticle production process in full phase
space in  $e^+e^-$ annihilation at the $Z_0$ peak is still lacking.
Experimental data show two interesting features:
a clear shoulder is visible in the intermediate multiplicity range in
charged particles multiplicity  distributions (MD's)\cite{delphi:2,opal,aleph};
the ratio of factorial cumulants over factorial
moments of the MD, $H_q$, when plotted as a function of its order $q$,
decreases sharply to a negative minimum at $q$=5 and follows  then a
quasi-oscillatory behavior\cite{sld,Gianini}.
In order
to interpret these results, two complementary approaches can be
followed.
First, properties of multiparton final states can be computed in the framework
of perturbative QCD by exploiting its branching structure\cite{KUV,DKMT};
partonic and hadronic distributions are then
directly compared by assuming the Local Parton Hadron Duality\cite{LPHD} or its
generalized version\cite{AGLVH:2} as hadronization prescription.
Second, a phenomenological approach can be taken and data compared with
simple parametrizations, like for instance
the Negative Binomial (NBD)\cite{AGLVH:1} or the  log-normal (LND)\cite{LND}
distributions.

Within the first approach, the partons' MD has been computed in Double Log
Approximation (DLA)\cite{DKMT}, getting a much wider MD than the
experimental one. By adding an approximate treatment of the energy-momentum
conservation law\cite{Dokshitzer}, the predicted MD becomes narrower at non
asymptotic energies, but it can
quantitatively reproduce neither the general shape of the experimental
MD nor its shoulder.
Concerning the ratio $H_q$, QCD predictions beyond DLA have been
considered\cite{DreminNech}. They qualitatively show the same behavior of
experimental data; the result supports the general picture of LPHD, but
it is quite far from being quantitatively satisfactory.

Within the second more phenomenological approach, let us consider
the two most popular parametrizations of MD's, the NBD and the LND.
Both of them are related to QCD. The NBD is the exact solution
for the  gluon MD in a quark-jet in Leading Log Approximation\cite{QCDAG}
and more detailed QCD calculations in DLA plus
recoil corrections turn out to be close to NB
behavior\cite{Dokshitzer,Mal-Web}. It reveals the self-similar and
Markoffian structure of parton shower evolution;
it is usually interpreted in the framework of the clan model\cite{AGLVH:1}
as an indication  of the two-step nature of the production process.
The LND results from the application of the central-limit theorem
to self-similar parton  showers\cite{LND}.
Both parametrizations quantitatively
describe  $e^+e^-$ annihilation data in full phase space at
c.m. energies below the $Z_0$ peak\cite{HRS,TASSO} and qualitatively reproduce
the overall shape of the MD at the $Z_0$ peak, but neither the NBD nor
the LND reproduces the shoulder structure, although the LND fares generally
better in terms of $\chi^2$/NDF.
Concerning the ratio $H_q$ of factorial cumulants over factorial
moments,  parametric models give a poor description of
experimental results. A full NBD predicts indeed positive definite and
monotonically
decreasing $H_q$'s. If one truncates a NBD  in the high multiplicity
tail to properly take into account the experimental cut due to finite
statistics, one recovers a quasi-oscillatory behavior qualitatively
consistent with experimental results\cite{hq}; the size of the experimental
effect is however strongly underestimated\cite{sld}.
Predictions of a truncated LND are in better agreement with data\cite{sld},
even though they do not provide a quantitative description of the
experimental effect.

In this letter, we propose
 a simple  phenomenological parametrization of the MD
in $e^+e^-$ annihilation in full phase space  in terms of a weighted
superposition of two NBD's; both the shoulder structure
of the MD and the quasi-oscillatory behavior of the ratio $H_q$ are
quantitatively reproduced. These results suggest to explain both effects
in terms of the superposition of
samples of events with a fixed number of jets, where  the standard NB
parametrization is indeed recovered.

\section{The shoulder structure}

The DELPHI Collaboration has shown in \cite{delphi:3} that
the shoulder structure in the MD in $e^+e^-$ annihilation can be explained
by  the superposition of the MD's coming from
events with 2, 3 and 4 jets, as identified by a suitable jet-finding algorithm,
and that the MD's in these classes of events are well reproduced by a single
NBD
for every value of the jet-finder parameter, $y_{cut}$.
It is thus suggested that the shoulder is associated with the radiation of hard
gluons resulting in the appearance of one of more extra jets in the hadronic
final state.

Let us also remind that a shoulder structure similar to the one observed in
$e^+e^-$ annihilation has been observed in $p\bar p$ collisions at high
energies\cite{UA5:3} and  was shown to be well described by
a 5-parameter parametrization in
terms of the weighted superposition of two  NBD's\cite{fuglesang}.

Inspired by these experimental results and NB universality for all
classes of reactions, we  propose now to parametrize
experimental data on MD's in full phase space in $e^+e^-$
annihilation at the $Z_0$ peak in terms of the superposition of 2 NBD's;
according to DELPHI's result, we argue that the two MD's should be
associated with the contribution of two- and multi($\ge$ 3)-jets
events respectively. We fix therefore the weight parameter
to be equal to the fraction of 2-jet events.
Formally, we perform a fit on available experimental data obtained at
LEP\cite{delphi:2,opal,aleph}\footnote{Data on MD by L3 Coll.\cite{l3} have
not been analyzed due to their large systematic error.}
 and SLC\cite{sld}
with the following 4-parameter distribution:
\be
P_n(\alpha ; \bar n_1, k_1; \bar n_2, k_2) = \alpha
P_n^{NB}(\bar n_1, k_1) + (1 - \alpha ) P_n^{NB}(\bar n_2, k_2)
\label{5par}
\ee
where $P_n^{NB}(\bar n,k)$
 is the NBD, expressed in terms of two parameters, the average multiplicity
$\bar n$
and the parameter $k$, linked to the dispersion by
$D^2/\bar n^2 = 1/\bar n + 1/k$,  as:
\be
P_n^{NB}(\bar n, k) = \frac{k(k+1)\dots (k+n-1)}{n!}
\left( \frac{k}{\bar n +k} \right)^k
 \left( \frac{\bar n}{\bar n + k} \right)^n
\ee
The weigth $\alpha$ gives the 2-jet events fraction:
since the latter depends on the jet-finder
parameter, $y_{cut}$,  different values of $\alpha$ corresponding to
different values of $y_{cut}$ have been considered.

As far as MD's in full phase space are concerned, one has also
to take care of the ``even-odd'' effect,
i.e., of the fact that the total numer of  final charged particles
must be even due to charge conservation;
accordingly, the actual form used in the fit procedure is given by:
\be
   P_n = \cases{ A P_n(\alpha ; \bar n_1, k_1; \bar n_2, k_2)
                    &if   $n$ \ even  \cr
                 0  & \hbox{otherwise} \cr}
\label{pnfps}
\ee
where $A$ is the normalization
parameter, so that $\sum_{n=0}^{\infty} P_n = 1$.

The parameters of the fits and the corresponding $\chi^2$/NDF for different
values of $\alpha$  are given in Table~1;
for comparison, the parameters of the NBD's fitted to 2- and 3-jet data samples
by DELPHI Coll.\cite{delphi:3} are also shown.
As an example, Figure~1 compares the experimental MD's
with the parametrization~\eref{pnfps} with best-fit parameters
given in Table~1 for $\alpha$ = 0.767;
the two NBD contributions are plotted with a dotted line.
The residuals, i.e., the
difference between data and theoretical predictions,
 are also shown below each MD in units of standard deviations.
 One concludes that
the proposed parametrization can reproduce  the experimental data very well;
no structure is visible in the residuals. The agreement holds for all
considered values of $\alpha$, even though the parameters of the two
NBD's obviously depend on the choice of the weight parameter.
To further support the interpretation of the two components of eq.~\eref{5par}
as the two- and multi-jet contribution respectively, we notice that the
parameters of the two NBD's  extracted
from the fit at a given $\alpha$
are close to the fitted values extracted
by DELPHI Coll. for 2- and 3-jet MD's at the corresponding
 value of $y_{cut}$;  in fact for these values of $y_{cut}$ the
contribution of $(\ge 4)$-jet events is negligible (less than 10\%).
Finally, let us also notice that the fit parameters present
large errors; their determination
would be strongly improved  by direct experimental analyses
which could take properly into account the full covariance matrix of the
MD's.  These analyses are eagerly awaited.

\section{The ratio $H_q$}

Let us consider now the ratio of factorial cumulant over
factorial moments
\be
  H_q = K_q / F_q                                  \label{hmom}
\ee
as a function of the order $q$.
The factorial moments, $F_q$, and
factorial cumulant moments, $K_q$,  can be obtained in general from the MD,
$P_n$, through the relations:
\be
  F_q = \sum_{n=q}^{\infty} n(n-1)\dots(n-q+1) P_n  ,         \label{facmom}
\ee
and
\be
  K_q = F_q - \sum_{i=1}^{q-1} {q-1 \choose i} K_{q-i} F_i .  \label{faccum}
\ee
A direct experimental analysis of the ratio $H_q$  with correlations and
systematics effects fully taken into account
has been so far performed by SLD Coll.\cite{sld} only.
To extend the analysis to LEP experiments, we  extracted {\it a posteriori}
the values of the $H_q$ from published MD's. Due to the lacking of the full
covariance matrix for MD's, the statistical error on $H_q$ cannot be properly
calculated; in order to  estimate it, for each experimental MD
we calculated the ratio $H_q$ from
1000 MD's obtained from the original one by
allowing gaussian fluctuations around the measured value of each $P_n$
the width of
the gaussian being given by the experimental statistical error.
 The standard deviation of the  $H_q$ distribution  for  each order $q$
 was then taken as an estimate of the
error associated with the ratio $H_q$ itself.
As a cross-check, this method has been applied
to SLD data, since both MD and the
ratio $H_q$ are available in this case;
a good agreement between the errors extracted with this procedure
and the sum of statistical and systematic
errors measured by SLD has been found. It should be stressed of course
 that this method can give
just an approximate value of the size of the errors
and direct experimental analyses by LEP Collaborations are awaited.
Let us now look at the ratio $H_q$ as predicted by the parametrization
proposed in Section~2.
 Since the $H_q$'s were shown to be sensitive to the truncation of the
tail due to the finite statistics of data samples\cite{hq}, moments were
extracted from a truncated MD defined as follows:
\be
   P_n^{trunc} = \cases{ A' P_n(\alpha; \bar n_1, k_1; \bar n_2, k_2)
                 &if ($n_{min} \le n \le n_{max}$)
                 \ \hbox{and} \ $n$ \ even  \cr
                 0  & \hbox{otherwise} \cr}
\label{pntrunc}
\ee
Here $n_{min}$ and $n_{max}$ are the minimum and  the maximum observed
multiplicity and $A'$ is a new normalization
parameter\footnote{We
found that a small (0.1\%) error in the normalization leads to very different
values for the $H_q$'s}, so that $\sum_{n=n_{min}}^{n_{max}} P_n^{trunc} = 1$.

Figure~2 shows the ratio $H_q$ as a function of the order $q$
for SLD\cite{sld}, ALEPH\cite{aleph},
DELPHI\cite{delphi:2} and OPAL\cite{opal}. It should be noticed that
SLD, DELPHI and OPAL give very similar results, whereas
ALEPH shows a different behavior, in particular
at low  orders of $q$ and after, when oscillations start.
However, due to the large errors and to the uncertainties in their estimate, a
definite answer on the consistency among different set of data requires a
 direct experimental determination of errors.
In  Figure~2 predictions of the  parametrization~\eref{pntrunc}
with parameters fitted to reproduce the MD as given in Table~1
for different values of $\alpha$ are also shown (solid lines).
It turns out that this  parametrization
reproduces not only the shoulder observed in experimental MD's
but  also quantitatively describes the experimental behavior of the ratio
$H_q$.
Results are essentially independent of  $\alpha$; the small
spread among theoretical predictions for different values of
$\alpha$ can be taken as an estimate of the theoretical error.
The plot referring to SLD  shows in addition predictions
of eq.~\eref{pnfps}, i.e., of the same parametrization, but without taking into
account the effect of truncation (dashed lines):
the quasi-oscillations become smaller.
To further support the interpretation of this oscillatory behavior as
due to the superposition of two- and multi-jet events, it should be
noticed that oscillations in samples of isolated 2- and 3-jet events are
much smaller than in the full sample of events\cite{Gianini}.
In conclusion, the observed behavior of $H_q$'s results  from the convolution
of two different effects, a physical one, i.e., the superposition of two
components, and a statistical one, i.e., the truncation of the tail due to the
finite statistics of  data samples.

\section{Conclusions}

Two experimental features of multiparticle production in full phase space in
$e^+e^-$ annihilation at the $Z_0$ peak, i.e., the shoulder in the
MD and
the oscillatory behavior of the ratio of factorial cumulants over factorial
moments, $H_q$, as a function of the order $q$, have been adressed:
a simple  phenomenological parametrization of the MD in terms
of a weighted superposition of two components,  each one
being distributed according to a  NBD, has been proposed.
The weight of the first component was
taken equal to the fraction of 2-jet events, i.e.,
the two components were identified with
the two- and the multi-jet contribution respectively.
The shoulder structure is found to be quantitatively reproduced by this
parametrization; the behavior of the ratio $H_q$ is also quantitatively
described, after taking properly into account the effect due to the truncation
of the tail. A consistent picture seems then to emerge:
both experimental effects which deviate by a single NBD behavior are
explained in terms of a common mechanism, linked to the emission of hard
gluons.
The simple NB parametrization
is reestablished at the level of events with fixed number of jets.
Further tests of the above mentioned picture can be provided
by direct experimental analyses which take into account the effects of
correlations and systematics; these analyses are eagerly awaited.

\section{Acknowledgements}

Useful discussions with G. Gianini, W. Ochs  and V. Uvarov are gratefully
acknowledged. We thank P. Burrows and J. Zhou for having provided us with
SLD data on MD.

\newpage

\section{Table Captions}

\begin{itemize}

\item[\bf Table 1.]
{Parameters and $\chi^2$/NDF of the fit to experimental data from
ALEPH\cite{aleph}, DELPHI\cite{delphi:2}, OPAL\cite{opal} and SLD\cite{sld}
 with the weighted sum of two NBD's.
Results are shown for different values of $\alpha$ corresponding to the
fraction of 2-jet events, $f$, experimentally measured by DELPHI Collaboration
  at different values of the jet-finder parameter
$y_{cut}$. NBD parameters extracted by the DELPHI Collaboration
by fitting MD's of samples of events with 2- and 3-jets
at different values of $y_{cut}$  are also shown for comparison in the last
columns.}

\end{itemize}

\section{Figure Captions}

\begin{itemize}

\item[\bf Figure 1.]
{Charged particles' MD in full phase space, $P_n$, at the $Z_0$ peak  from
ALEPH\cite{aleph}, DELPHI\cite{delphi:2}, SLD\cite{sld} and
OPAL\cite{opal} are
compared with equation~\eref{pnfps}
with $\alpha$ = 0.767 (see Table~1 for the values of the
corresponding parameters) (solid lines);
dotted lines indicate the two separate NBD contributions.
The lower part of the figure shows the residuals, $R_n$, i.e., the difference
between
data and theoretical predictions, expressed in units of standard deviations.}

\item[\bf Figure 2.]
{The ratio of factorial cumulnats over factorial moments,
$H_q$ as a function of $q$;
experimental data (diamonds) are compared  with equation~\eref{pntrunc}
for different values of $\alpha$ as in Table~1 (solid lines).
SLD, ALEPH, DELPHI and OPAL data are explicitly indicated.
The dashed lines in the first plot (SLD) show predictions of
equation~\eref{pnfps}, i.e., the same parametrization as the solid lines
 but without taking into account the effect of truncation.
In the figure only statistical errors of SLD data are shown.}

\end{itemize}

\newpage

\newcommand\pp{$\pm$}

\begin{table}
%\vspace{-3.0cm}
\centerline{Table 1}
\vspace{2.0cm}
 \begin{center}
 \vspace{4mm}
 \begin{tabular}{||c|c|c|c|c||c|c||}
\hline
& ALEPH\cite{aleph}  &   DELPHI\cite{delphi:2} &   OPAL\cite{opal}  &
SLD\cite{sld}
& \multicolumn{2}{c||}{~~~DELPHI\cite{delphi:3}}    \\ \hline
& \multicolumn{4}{l||}{~~$\alpha$ = 0.463}
& \multicolumn{2}{l||}{$y_{cut}$ = 0.02 \ \ \    $f$=0.463}  \\  \hline
$\bar n_1$ & 17.7\pp 1.1 &  18.2\pp 0.2     & 18.4\pp 0.2 & 18.4\pp 0.2 & $\bar
n_{2-jet}$ & 18.5\pp 0.1\\
$k_1$       & 111\pp 168 & 90\pp 20  & 71\pp 11 & 47\pp 4 & $k_{2-jet}$ & 57\pp
4 \\
$\bar n_2$  &     23.6\pp 0.8 & 23.9\pp 0.2  &   24.0\pp 0.2 &  23.0\pp 0.2 &
$\bar
n_{3-jet}$ & 22.9\pp 0.1 \\
$k_2$ &  32\pp 15 & 31\pp 3 &         28\pp 2 &   29\pp 2 &  $k_{3-jet}$
& 44\pp 2\\
$\chi^2$/NDF & 3.56/22 & 8.95/21 &  3.32/21 & 17.6/21 & & \\  \hline
& \multicolumn{4}{l||}{~~$\alpha$ = 0.659}  &
\multicolumn{2}{l||}{$y_{cut}$ = 0.04 \ \ \   $f$=0.659} \\  \hline
$\bar n_1$ & 18.5\pp 0.7 & 18.9\pp 0.2 & 19.0\pp 0.1 & 18.9\pp 0.1 & $\bar
n_{2-jet}$ &
19.4\pp 0.1\\
$k_1$       & 66\pp 46 & 63\pp 8  & 54\pp 5 & 42\pp 3 & $k_{2-jet}$ & 44\pp 2
\\
$\bar n_2$  &     25.5\pp 1.0 & 25.8\pp 0.3  &   25.9\pp 0.2 &  24.7\pp 0.2 &
$\bar
n_{3-jet}$ & 24.8\pp 0.1 \\
$k_2$  &   47\pp 33 &  44\pp 5 & 40\pp 5 &   37\pp 3 &  $k_{3-jet}$ & 42\pp 2\\
$\chi^2$/NDF & 3.72/22 & 10.1/21 &  4.40/21 & 16.3/21 & & \\  \hline
& \multicolumn{4}{l||}{~~$\alpha$ = 0.767}
& \multicolumn{2}{l||}{$y_{cut}$ = 0.06 \ \ \   $f$=0.767} \\  \hline
$\bar n_1$ & 19.1\pp 0.5 & 19.4\pp 0.2 & 19.5\pp 0.07 & 19.3\pp 0.09 & $\bar
n_{2-jet}$ &
20.0\pp 0.1\\
$k_1$       & 53\pp 24 & 52\pp 6  & 46\pp 3 & 39\pp 2 & $k_{2-jet}$ & 38\pp 1
\\
$\bar n_2$  &     27.0\pp 1.1 & 27.3\pp 0.3  &   27.5\pp 0.2 &  26.0\pp 0.2 &
$\bar
n_{3-jet}$ & 26.0\pp 0.1 \\
$k_2$  &   65\pp 62 &  61\pp 10 & 55\pp 8 &   47\pp 5 &  $k_{3-jet}$ & 45\pp
2\\
$\chi^2$/NDF & 3.86/22 & 11.7/21 &  6.30/21 & 15.6/21 & & \\  \hline
& \multicolumn{4}{l||}{~~$\alpha$ = 0.834}
& \multicolumn{2}{l||}{$y_{cut}$ = 0.08 \ \ \  $f$=0.834} \\  \hline
$\bar n_1$ & 19.5\pp 0.4  &  19.8\pp 0.1   & 19.9\pp 0.6 & 19.6\pp 0.1 & $\bar
n_{2-jet}$ & 20.4\pp 0.1\\
$k_1$       & 45\pp 15 & 46\pp 3  & 40\pp 2 & 37\pp 2 & $k_{2-jet}$ & 34\pp 1
\\
$\bar n_2$  &  28.2\pp 1.2  & 28.6\pp 0.3  &  28.8\pp 0.2 &  27.1\pp 0.3 &
$\bar
n_{3-jet}$ & 26.8\pp 0.1 \\
$k_2$  &   92\pp 121 &  85\pp 18 &  76\pp 15 &   59\pp 7 &  $k_{3-jet}$ & 49\pp
1 \\
$\chi^2$/NDF & 3.99/22 & 13.9/21 &  8.81/21 &  15.2/21 & & \\  \hline
 \end{tabular}
 \end{center}
\label{table}
\end{table}

\newpage

\begin{figure}
          \begin{center}
\mbox{\epsfig{file=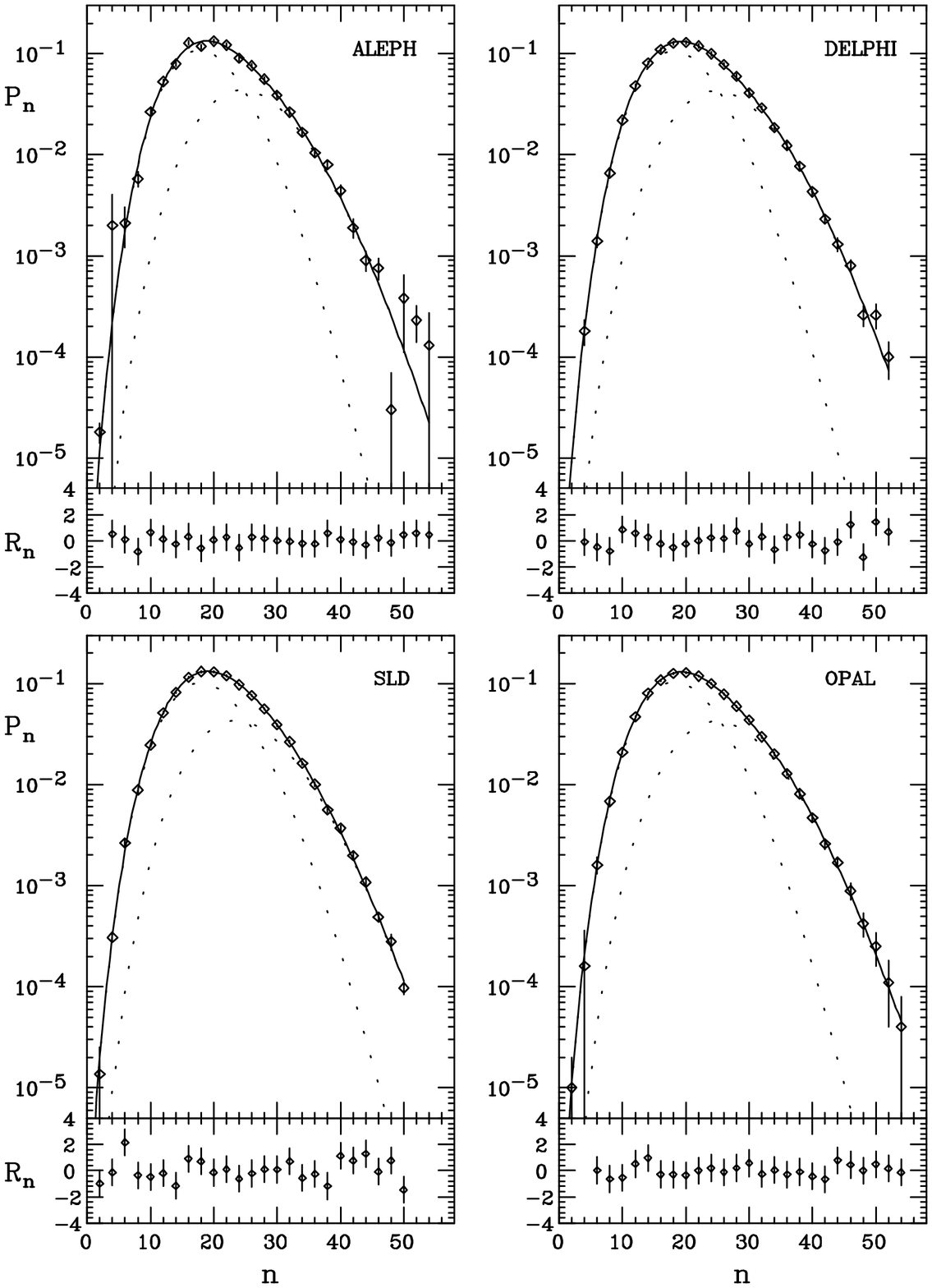,bbllx=3.cm,bblly=2.cm,bburx=19cm,%
bbury=26.cm,width=15cm}}
\end{center}
\caption{}
\end{figure}

\newpage

\begin{figure}
          \begin{center}
\mbox{\epsfig{file=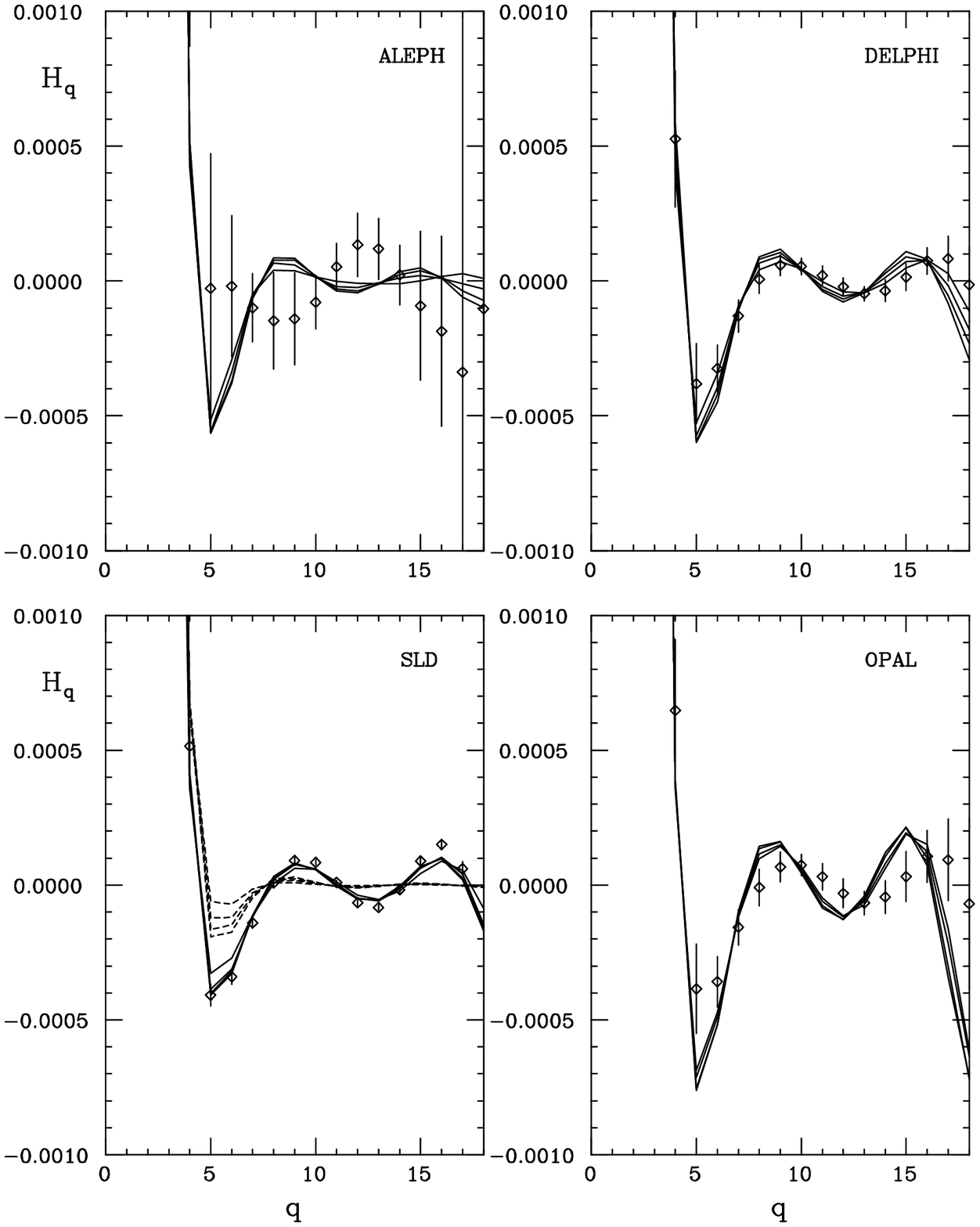,bbllx=2.cm,bblly=4.cm,bburx=19cm,%
bbury=24.cm,width=17cm}}
\end{center}
\caption{}
\end{figure}

\end{document}